\documentclass[runningheads]{llncs}
\usepackage[T1]{fontenc}
\usepackage{graphicx}
\usepackage{comment}
\usepackage{float}
\usepackage{amsmath}
\usepackage{algorithm}
\usepackage{algpseudocode}
\usepackage{stmaryrd} 
\usepackage{seqsplit}
\usepackage{subcaption}

\usepackage{xcolor}
\usepackage{todonotes}

\definecolor{atomictangerine}{rgb}{1.0, 0.6, 0.4}

\definecolor{brightube}{rgb}{0.82, 0.62, 0.91}

\definecolor{brightturquoise}{rgb}{0.03, 0.91, 0.87}

\newcommand{\bsq}[1]{\lq{#1}\rq}
\usepackage{hyperref}
\usepackage{color}

\DeclareMathOperator{\opnot}{not}
\DeclareMathOperator{\nand}{nand}

\usepackage{siunitx}
\newcommand{\stitch}{\textsc{Stitch}}
\newcommand{\augment}{\oplus}
\newcommand{\terms}[1]{T(#1)}

\begin{document}
\title{Twitch: Learning Abstractions for Equational Theorem Proving}
%
%
\author{Guy Axelrod\inst{1,2}\orcidID{0000-0002-1752-8069} \and
Moa Johansson\inst{1,2}\orcidID{0000-0002-1097-8278} \and
Nicholas Smallbone\inst{1,2}\orcidID{0000-0003-2880-6121}}
\authorrunning{G. Axelrod et al.}
%
\institute{Chalmers University, of Technology, Gothenburg, Sweden \and
University of Gothenburg, Gothenburg, Sweden
\email{\{guya,jomoa,nicsma\}@chalmers.se}}
\maketitle              
%


\begin{abstract}
  Several successful strategies in automated reasoning rely on human-supplied guidance about which term or clause shapes are interesting. In this paper we aim to discover interesting term shapes \emph{automatically}. Specifically, we discover \emph{abstractions}: term patterns that occur over and over again in relevant proofs. We present our tool \emph{Twitch} which discovers abstractions with the help of \stitch, a tool originally developed for discovering reusable library functions in program synthesis tasks. Twitch can produce abstractions in two ways: (1) from a partial, failed proof of a conjecture; (2) from successful proofs of other theorems in the same domain. We have also extended Twee, an equational theorem prover, to use these abstractions.
  We evaluate Twitch on a set of unit equality (UEQ) problems from TPTP, and show that it can prove 12 rating-1 problems as well as yielding significant speed-ups on many other problems.

\end{abstract}
\section{Introduction}

Automated theorem provers must navigate a massive search space, choosing the useful inferences and discarding the useless ones. For example, a proof in Twee \cite{smallbone2021twee} may consider hundreds of millions of critical pairs, selecting some tens of thousands as active rewrite rules, of which perhaps a few hundred are used in the final proof. Thus, it is essential to identify which critical pairs are interesting -- but Twee's heuristics for this are simplistic, relying mostly on the size of the critical pair. It is hard to do anything smarter because, with so many possible critical pairs, each one must be judged quickly.

This story is typical for saturation-based theorem provers: very many possible inferences, ranked using a simple metric. This is quite effective, but sometimes more intelligence is needed. One approach popularised by Otter \cite{otter} is to inject human guidance about which \emph{shapes} of terms or clauses are interesting. The user specifies a set of interesting shapes, and generated clauses that match these shapes are preferred over ones that do not. This idea originated in the \emph{weighting strategy} \cite{weighting} and was later refined by the \emph{resonance strategy} \cite{wos1995,beeson2016,fitelson2001} and \emph{hint strategy} \cite{veroff1996,veroff2001}. All three strategies have been used to find difficult proofs.

In this paper we find interesting term shapes \emph{automatically}. We analyse failed proof attempts and proofs of related conjectures to find term shapes which occur often, which we call \emph{abstractions}. We have added support for abstractions to Twee, and with the help of automatically generated abstractions we are able to prove 12 rating-1 problems from the TPTP \cite{sutcliffe1998tptp}, as well as several other problems that Twee previously could not solve (see Section~\ref{sec:experiments}). Our contributions are:

\begin{itemize}
\item A method for automatically generating abstractions from an \emph{unsuccessful attempt} to prove a conjecture (see Section~\ref{sec:partial_proof_abstractions}). The idea is to pick out interesting lemmas from the proof attempt, and find patterns of terms that occur repeatedly in the proofs of those lemmas. To find the patterns we use the tool \stitch{} \cite{bowers2023}, which attempts to \emph{compress} a set of terms (here from proofs) by introducing auxiliary function definitions as abbreviations.

\item A method for generating abstractions from a corpus of successful proofs in the same domain, again using \stitch{} (see Section~\ref{sec:domain_abstractions}). The hypothesis is that patterns of terms that occur repeatedly in the proofs are often good abstractions for other conjectures in the same domain.
  The aim here is to create a methodology inspired by curriculum learning \cite{curriculum-learning} for proving difficult equational problems: first prove some simpler problems in the same domain, abstract out common patterns appearing in those proofs, and use those abstractions to guide the proof of the difficult problem.

\item An implementation of abstractions in the equational theorem prover Twee \cite{smallbone2021twee}, inspired by the weighting strategy \cite{weighting} (see Section~\ref{sec:hints_in_twee}).
\end{itemize}

The code for Twitch as well as all data and experimental results are supplied in the associated repository \cite{repository}.


\subsection{Motivating Example}
\label{sec:example}

Let us consider the problem \texttt{LAT075-1} from the TPTP, which concerns the equational axiomatization of modular ortholattices in a language with a single function symbol $f$ representing the Sheffer stroke (or NAND operator). The details of this problem are not important here, but what is relevant is that the same kind of term appears repeatedly in the proof that Twee finds. Here is a short fragment of the proof:


{
\allowdisplaybreaks
\begin{flalign*}
&\vdots\\
&= f(x, f(y, \boxed{f(f(z, {\boxed{f(x, x)}}), f(z, {\boxed{f(x, x)}}))}))\\
&= f(x, f(y, f(f(z, \boxed{f(x, x)}), f(\boxed{f(x, x)}, z))))\\
&= f(x, f(y, \boxed{f(f({\boxed{f(x, x)}}, z), f({\boxed{f(x, x)}}, z))}))\\
&= f(\boxed{f({\boxed{f(x, x)}}, {\boxed{f(x, x)}})}, f(y, \boxed{f(f({\boxed{f(x, x)}}, z), f({\boxed{f(x, x)}}, z))}))\\
&\vdots
\end{flalign*}
}


As we can see, terms of the form $f(t, t)$, which we have marked with a box above, appear over and over in the proof.\footnote{This is perhaps not so surprising as $f(t,t)$ represents the negation of $t$.}
Our hypothesis is that \emph{the same term shape} is likely to be common in proofs of other theorems about the Sheffer stroke.


This gives us the following basic approach to proving a difficult theorem. First we analyse a proof of a related theorem to find commonly occurring term patterns. Then, on the assumption that these term patterns should also occur in the proof of the difficult theorem, we ask the theorem prover to prefer inferences involving those term patterns. The remaining questions are, how can we find these term patterns, and how can the prover exploit them?

To formalize the notion of \emph{common patterns} in a proof, we imagine \emph{compressing} the proof by introducing new definitions to abbreviate certain terms. For example, if we were to define $g(x) := f(x, x)$, then the proof above could be written more compactly; e.g. the first term could be written as $f(x, f(y, g(f(z, g(x)))))$. Then we ask: what is a \emph{maximally compressive} set of abbreviations, that is, a set of definitions that allows us to express the proof terms as compactly as possible?


The tool \stitch{} \cite{bowers2023} (see Section~\ref{sec:stitch}) is designed to answer exactly this question. Given a set of terms, it searches for the maximally compressive function definitions. Henceforth, we will call such function definitions \emph{abstractions}. In this case, \stitch{} would find that $g(x) := f(x, x)$ is a maximally compressive abstraction (and when given the full proof, it also finds other useful abstractions, such as the AND operator $h(x,y) := f(f(x,y),f(x,y))$).

Once we have these abstractions, how can we use them to influence proof search? One possibility is to simply add the function definition as an axiom, in this case $\forall x.\, g(x) = f(x,x)$. This would allow the prover to rewrite terms of the form $f(t, t)$ to $g(t)$, which has a reduced cost as it is smaller, so that if such a term occurs in a critical pair then the critical pair is more likely to be selected. For \texttt{LAT075-1} this approach works well, and reduces the runtime of Twee from $\sim 250 s$ to $\sim 10 s$.

Adding definitional axioms that abbreviate common term patterns is a known technique that often works well.\footnote{For example, see the Robbins problem variant \texttt{ROB033-1} in the TPTP. It is also the basis of Twee's goal direction heuristic \cite{smallbone2021twee}.} However, adding new axioms also allows new inferences between them, thus expanding the search space. As we show in Section~\ref{sec:ablations}, adding more than a few such axioms harms the prover's performance.



Instead, we have extended Twee to natively support abstractions (see Section~\ref{sec:hints_in_twee}). The abstractions influence the heuristic function of the prover, so that inferences matching them are more likely to be selected, but they do not alter the search space itself. Because of this, we can provide many more abstractions, but the individual abstractions may have a smaller effect. In this case, passing $f(x, x)$ as an abstraction reduces the runtime of Twee from $\sim 250 s$ to $\sim 130 s$.

\section{Background: \stitch}
\label{sec:stitch}

\stitch{} \cite{bowers2023} is a system originally designed for library learning in program synthesis -- automatically discovering reusable function abstractions from an existing collection of programs. Given a set of terms as input, it constructs functions which generalize parts of the terms, and uses those functions to abbreviate the terms. \stitch{} attempts to construct \emph{maximally compressive} abstractions; specifically, it maximizes the product of the size of the function and the number of places where the function can be used.

Although \stitch{} was designed to work on program code, it accepts as input any set of terms. Hence, given a Twee proof, we can give \stitch{} the set of all terms occurring in the proof and it will find abstractions for it.

\label{sec:translation}

\stitch's input language is a higher-order $\lambda$-calculus. Thus, we need to translate proof terms into $\lambda$-terms, which we do by treating function symbols as constants, and $\lambda$-abstracting all free variables. For example, the term $f(x,\; f(y,\; f(f(z,f(x,x)),\; f(z,f(x,x)))))$ from our example is translated into $\lambda x.\lambda y.\lambda z.\; (f\ x\ (f\ y\ (f\ (f\ z\ (f\ x\ x))\ (f\ z\ (f\ x\ x)))))$.
%
%
In this case, \stitch{} will return the
abstraction $g(\alpha) := (f\ \alpha\ \alpha)$,
where $\alpha$ is a \emph{schematic parameter} standing for an arbitrary subterm, which can be used to compress the input term to $\lambda x.\lambda y.\lambda z.\; (f\ x\ (f\ y\ g(f\ z g(x))))$.
%
Because the body of $g$ has a first-order shape, it can be translated back
directly, yielding the abstraction $f(x,x)$.

In general, \stitch{} abstractions are arbitrary $\lambda$-terms and can be higher-order. In practice, \stitch{} rarely generates higher-order abstractions in our setting since the input terms are first-order. If it does, we discard the abstraction.

\section{Twee + \stitch{} = Twitch}
\label{sec:twitch}
\begin{figure}[htbp]
    \centering
\includegraphics[scale=0.5]{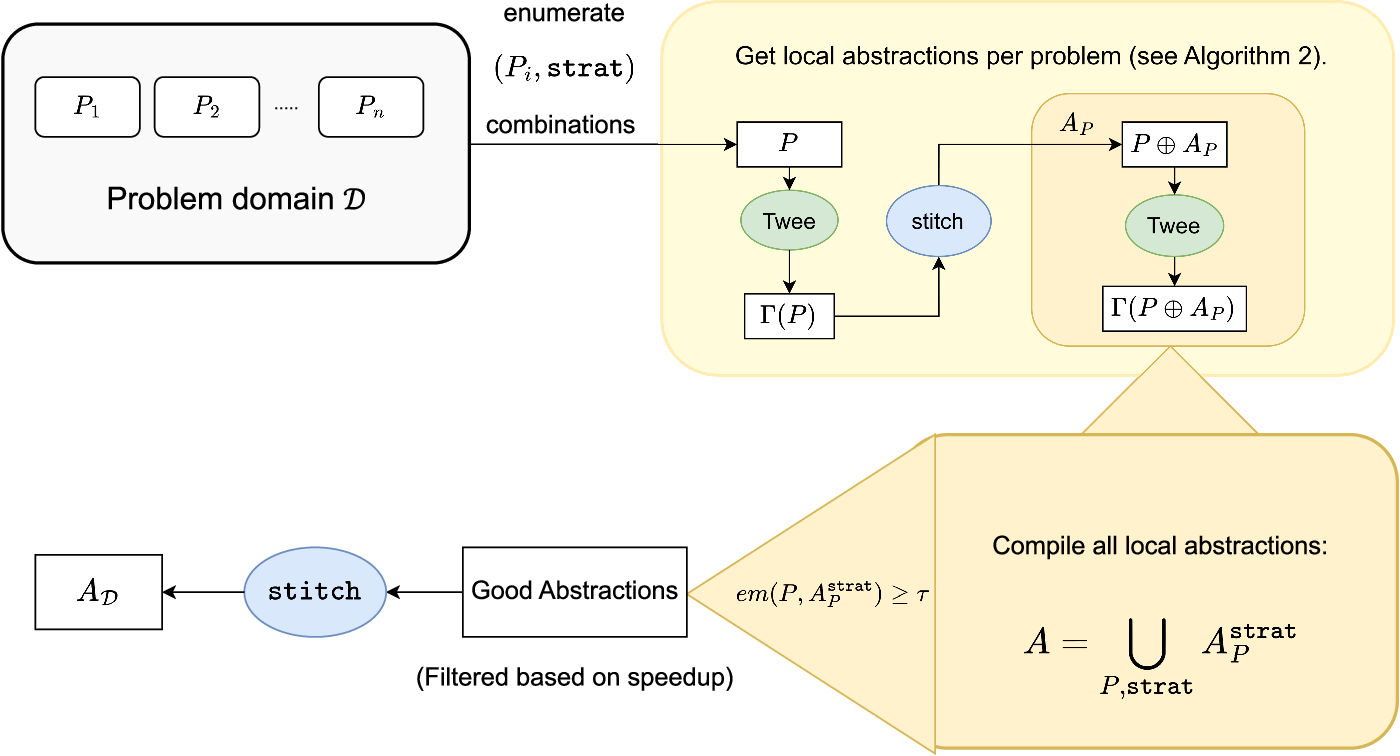}
    \caption{Overview of domain abstraction generation via Twitch (Algorithm~\ref{alg:get_domain_abstractions} and its subroutine Algorithm~\ref{alg:local_abstractions}).}
    \label{fig:twitch}
\end{figure}

In this section we present the implementation of Twitch: a system for automating the discovery of reusable abstractions that can be used to help solve hard problems in a given domain. Twitch has two modes of usage: abstractions can be discovered from partial proof attempts as described in section \ref{sec:partial_proof_abstractions}, or from easier proofs in the same domain, as described in section \ref{sec:domain_abstractions}. We say that the first mode produces \emph{partial proof abstractions}, while the second produces \emph{domain abstractions}. An overview of how domain abstractions are derived is shown in Figure \ref{fig:twitch}. The abstractions found by Twitch are used to adjust the search heuristics in the Twee prover, as described in Section~\ref{sec:hints_in_twee}.

Henceforth, we denote a given input problem -- which consists of a set of equations (axioms) and a goal/conjecture equation to prove -- by $P$.

\subsection{Abstractions from Partial Proofs}
\label{sec:partial_proof_abstractions}
The first approach we explore discovers abstractions given as input a single hard problem $P_h$, and can be used when we do not have examples of related problems to learn from. We start by running Twee (without abstractions) on $P_h$ for a fixed amount of time. When it times out, even though it has not proved the goal, it will have derived a large collection of rewrite rules. Each of these derived rules $l \to r$ corresponds to an equation $l = r$ about the theory we are exploring. Many of these equations will be quite boring, but some will represent interesting lemmas. Our idea is to learn abstractions from only the most interesting lemmas.

In order to define interestingness, we first need some auxiliary definitions. For any lemma $l = r$, we shall denote by $T(l = r)$ the set of terms occurring in its proof, including the proofs of any sub-lemmas used in the proof. More precisely, the proof of $l = r$ is a chain $l = t_1 = \ldots = t_n = r$, where each step may be justified by a lemma $l_i=r_i$ having its own proof. We then define $T(l = r) = \left\{l, t_1, \ldots, t_n, r\right\} \cup \bigcup_i T(l_i=r_i)$.


For a term $t$, we define its weight $|t|$ as the number of function applications plus the number of unique variables occurring in $t$. We extend this to equations by defining $|l=r| = |l| + |r|$, and to sets of terms $T$ by defining $|T| = \sum_{t \in T} |t|$. We then define the ``interestingness'' $s(l = r)$ of a lemma as follows:
$$s(l = r) := \frac{|T(l=r)|}{|l=r|^2}.$$
The intuition is that lemmas with higher scores are those that are simple statements requiring a long proof. Of course this does not necessarily coincide with what a human would consider interesting or useful, but we take it as a rough heuristic towards measuring these notions.

Given some partial proof of $P_h$, we sort the lemmas discovered in decreasing order of $s$. We then extract the first $k$ lemmas $l_1 = r_1, \ldots, l_k = r_k$. What we call the \emph{partial proof abstractions} are simply those we get by running \stitch{} on $\bigcup_{1\leq i\leq k} T(l_i = r_i)$.
We can then rerun Twee on $P_h$ augmented with these abstractions.


\subsection{Domain Abstractions}
\label{sec:domain_abstractions}
The goal of this approach is to learn abstractions that are useful for proving problems in a particular domain $\mathcal{D}$. We require a set of problems, some of which are already proved (the ``easy'' problems) and some of which are not (the ``hard'' problems). 
The algorithm works as follows:
\begin{enumerate}
\item First, we use \stitch{} on the proofs of the easy problems to find abstractions for each problem. These abstractions are \emph{local}, i.e. specific to one problem.

\item Then, we augment each problem with its local abstractions and re-run Twee to empirically establish if the abstractions are \emph{good}, i.e. if they allow us to reprove the problem faster.

\item Finally, we build a set of all the good local abstractions, and compress it with \stitch{} to find commonalities and derive higher-level \emph{domain abstractions}.
\end{enumerate}
This process is illustrated in Figure~\ref{fig:twitch}, with pseudocode in Algorithm~\ref{alg:get_domain_abstractions}.  There, we use $\terms{\Gamma(P)}$ to denote the set of terms occurring in the proof $\Gamma(P)$ of problem $P$ found by Twee. The set of abstractions derived by running \stitch{} on $\terms{\Gamma(P)}$ (where $\Gamma(P)$ was found using a specific Twee strategy \texttt{strat}) is denoted by $A^{\texttt{strat}}_P$. We use $P \augment A$ to denote $P$ augmented with abstractions $A$. Finally, the local abstractions are filtered by an effect metric $em$ which measures the speedup obtained by augmenting the problem with the abstraction. More precisely, 
$$em(P, A) = \frac{t(\Gamma(P))}{t(\Gamma(P \augment A))},$$
where $t(\Gamma(P))$ denotes the time taken by Twee to find proof $\Gamma(P)$.
Only those abstractions that lead to a speedup above some threshold $\tau$ are added to the set of good abstractions.

\begin{algorithm}[bp]
\caption{}
\label{alg:get_domain_abstractions}
\begin{algorithmic}[1]
\Procedure{DomainAbstractions}{$\mathcal{D}$}
    \State $\texttt{good\_absts} \gets$ empty set
    \ForAll{combinations of $P \in \mathcal{D}$ and $\texttt{strat} \in \texttt{strategies}$}
        \State $A^{\texttt{strat}}_P , em(P, A^{\texttt{strat}}_P) \gets$ \textsc{LocalAbstractions}($P$, \texttt{strat})
        \If{$em(P, A^{\texttt{strat}}_P) \geq \tau$}
            \State Add the alpha-normalized $A^{\texttt{strat}}_P$ to \texttt{good\_absts}
        \EndIf
    \EndFor
    \State $\mathcal{D} \gets$ Run \texttt{\stitch} over \texttt{good\_absts}
    \State Return $\mathcal{D}$
\EndProcedure
\end{algorithmic}
\end{algorithm}

\begin{algorithm}[tbp]
\caption{}
\label{alg:local_abstractions}
\begin{algorithmic}[1]
\Procedure{LocalAbstractions}{$P$, \texttt{strat}}
  \State Run Twee on $P$ with \texttt{strat}. 
  \State If it fails to terminate within the time limit, return failure.
  \State Else we get proof $\Gamma(P)$.
  \State Extract terms from $\Gamma(P)$ as strings to get list $\terms{\Gamma(P)}$.
  \State Run \stitch{} on $\terms{\Gamma(P)}$ to get abstractions $A^{\texttt{strat}}_P$.\label{line:run_stitch}
  \State Run Twee on $P \augment A^{\texttt{strat}}_P$ with \texttt{strat}.\label{line:augmented_run}
 \State Compute $em(P, A^{\texttt{strat}}_P)$. If Twee timed out on $P \augment A^{\texttt{strat}}_P$, set $em(P, A^{\texttt{strat}}_P) = 0$.
 \State Return $A^{\texttt{strat}}_P$, $em(P, A^{\texttt{strat}}_P)$.
\EndProcedure
\end{algorithmic}
\end{algorithm}

We see that growing the good abstraction set relies on running the \textsc{LocalAbstractions} subroutine for a collection of different \emph{strategies}. This simply refers to an assignment of specific values to parameters that influence the working of Twee. In particular, the strategy parameters are (1) whether to run Twee with or without goal flattening and (2) the abstraction weight factor (see Section~\ref{sec:hint_cost}). By iterating over combinations of these parameters, we increase the pool of good abstractions from which to derive domain abstractions later.

We can then use these domain abstractions to augment hard problems in the same domain (see Section \ref{sec:hard_problems}). In practice, given a hard problem $P_h \in \mathcal{D}$, we perform a filtering step on the good abstractions before running \stitch{} for the second time (i.e. before line 10 in Algorithm~\ref{alg:get_domain_abstractions}). Specifically, we remove any abstraction that contain function symbols not present in $P_h$, or shared function symbols with mismatching arities. This ensures that the resulting domain abstractions are compatible with $P_h$.

\section{Supporting Abstractions in Twee}
\label{sec:hints_in_twee}

Twee's architecture is based on \emph{unfailing completion} \cite{completion,unfailing-completion}, which works by treating the input equations as rewrite rules and then deriving new rules that cannot be proved by forward rewriting. For example, given some axioms about group theory, Twee might orient them into the following rules:
\[
e\cdot x \rightarrow x\qquad
x^{-1}\cdot x \rightarrow e\qquad
(x\cdot y)\cdot z \rightarrow x\cdot(y\cdot z)
\]

We can observe that the term $(x^{-1}\cdot x)\cdot y$ has two normal forms:
\[
(x^{-1}\cdot x)\cdot y \rightarrow e \cdot y \rightarrow y \qquad
(x^{-1}\cdot x)\cdot y \rightarrow x^{-1} \cdot (x \cdot y),
\]
hence the equation $x^{-1} \cdot (x \cdot y) = y$ is true but cannot be proved by forward rewriting, since both sides are normal forms (this is called a \emph{critical pair}). Twee therefore adds the new rewrite rule $x^{-1} \cdot (x \cdot y) \rightarrow y$. This process of deriving new rewrite rules continues until the rule set is saturated or the goal is proved.

One problem is that the number of critical pairs typically grows quadratically in the number of derived rules: a typical proof may produce $\sim 10^5$ rules with $\sim 10^{10}$ possible critical pairs. Hence, we cannot turn \emph{all} critical pairs into rules. Twee instead uses a heuristic approach where, at each step, the ``best'' critical pair is turned into a rewrite rule. Twee uses several criteria to rank the critical pairs (see \cite{smallbone2021twee}) but the most important one is the equation's \emph{weight}, defined as:

\begin{align*}
    w(t = u) &= w(t) + w(u) \\
    w(x) &= 1, \textrm{ if $x$ is a variable} \\
    w(F(t_1, \ldots, t_n)) &= 1 + \sum_{i=1}^n w(t_i), \textrm{ if $F$ is a function symbol }
\end{align*}

\newcommand{\fplus}{{\operatorname{\mathit{plus}}}}
\newcommand{\ftimes}{{\operatorname{\mathit{times}}}}

Our approach to abstractions is inspired by the \emph{weighting strategy} \cite{weighting}: whenever a term or subterm in a critical pair matches an abstraction, reduce its computed weight. Specifically, we extend $w$ with the following clause:
\begin{align*}
    w(A\sigma) &= w_A + \sum_{x \in dom(\sigma)} w(x\sigma), \textrm{ if $A$ is an abstraction and $\sigma$ is a substitution}
\end{align*}
Here, $w_A$ is a constant representing the weight of the abstraction. We come back to this in Section~\ref{sec:hint_cost}, but for now, let us assume that $w_A = 1$.

For example, suppose we are studying two functions $\ftimes$ and $\fplus$.
To encourage the prover to reason about distributivity, we might give the following abstraction $A$, representing a term in which distributivity can be applied.
%
\[
   \fplus(\ftimes(x, y), \ftimes(x, z)) \label{eq:hint}
\]
Now consider the following equation, perhaps occurring as a critical pair:
\[
\ftimes(c, \underline{\fplus(\ftimes(g(f(a),b), f(x)), \ftimes(g(f(a),b), g(y,z)))}) = g(a,b)
\]
Without the abstraction $A$,
this equation would have a weight of 21. With $A$, the underlined term can be matched as $A\sigma$,  where $\sigma = \{x \mapsto g(f(a),b),$ $y \mapsto f(x), z \mapsto g(y, z)\}$. Hence the total weight is $1 + w(c) + w(A\sigma) + w(g(a,b)) = 2 + w_A + w(g(f(a),b)) + w(f(x)) + w(g(y, z)) + 3 = 15$. This means that, if this equation is produced as a critical pair, it is more likely to be selected.

Note that using $A$ reduces the cost of the equation above in two ways: (1) the ``skeleton'' of the abstraction,  $\fplus(\ftimes(\_, \_), \ftimes(\_, \_))$, only counts for weight 1 instead of 3; (2) although $x$ occurs twice in the abstraction, $x\sigma$ is only counted once, hence the repeated subterm $g(f(a),b)$ is only counted once. Property (2) is vital in order for abstractions to act like \emph{abbreviations}. For example, in Section~\ref{sec:example} we define the abstraction $f(x,x)$, where $f$ is the Sheffer stroke, which represents the negation of $x$. Hence we should like $f(t,t)$ to have the same weight as a hypothetical term $not(t)$, which requires property (2).

As an aside, we have also implemented term-level \emph{resonators} \cite{wos1995} in Twee. These work just the same as abstractions, except that they only match if every variable in $\sigma$ is mapped to a variable (not a compound term).

\subsection{The Cost of an Abstraction}
\label{sec:hint_cost}

In the example above, we assumed that the weight $w_A$ of an abstraction was 1 symbol. In reality, we allow the user to choose between two methods of computing the weight. Both methods are parametrized on a constant $k$ chosen by the user:

\begin{description}
\item[Constant weight] 
Each abstraction $A$ gets a fixed weight $w_A = k$.

\item[Weight factor] 
This method gives larger abstractions greater weight.
Given the abstraction $A$, we first compute the weight of $A$'s \emph{skeleton}, $w(skel(A))$. This is the weight of $A$ if all variables are counted as weight 0 (i.e. ignored). In the example above, $w(skel(A)) = 3$.
Then we set $w_A = w(skel(A)) \times k$.
\end{description}

\section{TPTP Experiments}
\label{sec:experiments}


We have evaluated Twitch on a set of unsatisfiable unit equality (UEQ) problems from TPTP v9.2.1 \cite{sutcliffe1998tptp}. The problems are drawn from nine TPTP domains, summarized in Table~\ref{tab:tptpv9}. The main purpose is to measure whether the abstractions discovered by Twitch help in solving problems from the TPTP.

Section~\ref{sec:timeslice} measures whether partial proof abstractions and domain abstractions improve Twee's runtime. For domain abstractions, the user must define which problems make up the domain; in our experiments (with one exception mentioned in Section~\ref{sec:domain_hard_problems_solved}), we simply considered the domain to be the set of problems in a given TPTP domain (e.g. Group Theory (GRP), Lattice Theory (LAT), etc.) 

Section~\ref{sec:hard_problems} reports on the hard problems that abstractions from Twitch allow us to solve. Here, we define a problem as hard if it has TPTP rating at least 0.9 (i.e. 90\% of ATPs cannot solve it within 300 seconds) and Twee (without abstractions) cannot solve it within 1000 seconds. There are 70 such problems from Table~\ref{tab:tptpv9}.

Section~\ref{sec:ablations} considers various ablations testing some aspects of Twitch's architecture and strategies.

For details regarding relevant parameter values used, see Appendix~\ref{sec:parameters}.


\begin{table}[b]
    \centering
    \begin{tabular}{lr}
	\textbf{Domain} & \textbf{Count} \\
      \hline
      ALG (General Algebra) & 20 \\
      BOO (Boolean Algebra) & 56 \\
      COL (Combinatory Logic) & 120 \\
      GRP (Group Theory) & 481 \\
      LAT (Lattice Theory) & 126 \\
      LCL (Logic Calculi) & 84 \\
      REL (Relation Algebra) & 81 \\
      RNG (Ring Theory) & 48 \\
      ROB (Robbins Algebra) & 25 \\
      \hline
      Total & 1041 \\
    \end{tabular}
    \caption{Unsatisfiable UEQ problems per TPTP domain, TPTP v9.2.1}
    \label{tab:tptpv9}
\end{table}


\subsection{Overall Runtime Improvements}
\label{sec:timeslice}
\begin{figure}[htbp]
    \centering
    \begin{subfigure}{0.49\textwidth}
        \centering
        \includegraphics[width=\linewidth]{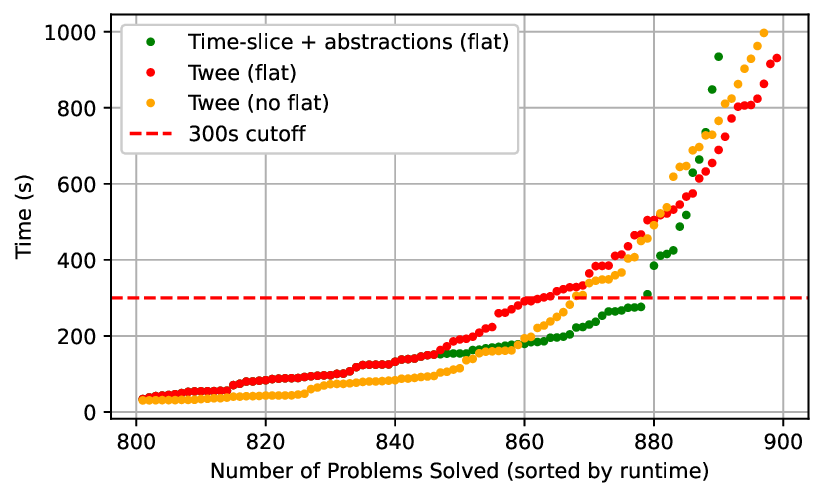}
        \caption{Using partial proof abstractions.}
        \label{fig:partial-timeslice}
    \end{subfigure}
    \hfill
    \begin{subfigure}{0.49\textwidth}
        \centering
        \includegraphics[width=\linewidth]{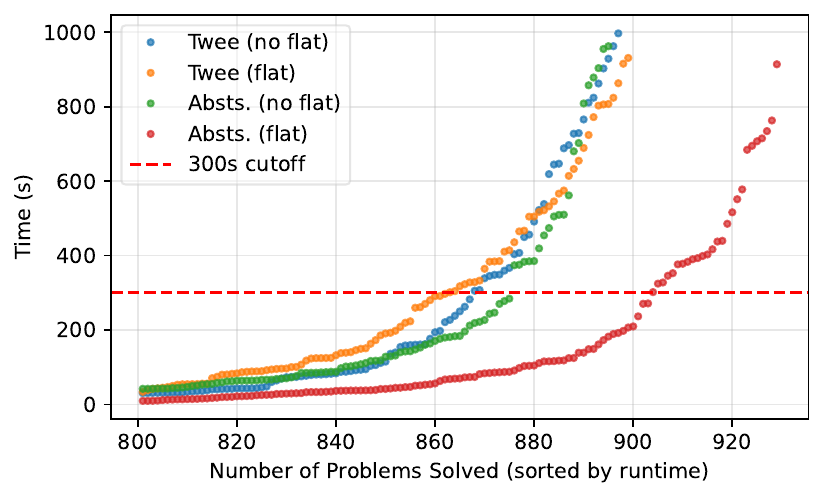}
        \caption{Using domain abstractions.}
        \label{fig:domain-timeslice}
    \end{subfigure}
    \caption{Cactus plots of Twee runtimes for TPTP problems. A given point in these plots corresponds to a particular problem $P$ from Table \ref{tab:tptpv9}. The y-value of a point indicates the time taken by Twee to find a proof of $P$ (possibly augmented with Twitch domain abstractions), and the corresponding x-value is the index of $P$ when the problems are sorted in increasing order of runtime.}
\end{figure}

Figures \ref{fig:partial-timeslice} and \ref{fig:domain-timeslice} include cactus plots of runtimes of Twee on the TPTP problems listed in Table~\ref{tab:tptpv9}. We benchmark Twee both with and without its \emph{goal flattening} strategy \cite{smallbone2021twee}, in order to see how this strategy interacts with abstractions. In both figures, the \bsq{Twee (flat)} and \bsq{Twee (no flat)} plots show Twee running without abstractions (with and without goal flattening respectively) for the full time limit of 1000s. We have plotted a line at 300s, a typical timeout for ATP benchmarking, but continue the evaluation until 1000s.

Figure~\ref{fig:partial-timeslice} shows the results for partial proof abstractions. Here, we first run Twee (without goal flattening) for 150 seconds, then if it times out, extract partial proof abstractions from the failed proof, and rerun Twee (with goal flattening) augmented with these abstractions for the remaining time.

Figure~\ref{fig:domain-timeslice} shows the results when using domain abstractions. As before, the \bsq{Twee (flat)} and \bsq{Twee (no flat)} show the performance of Twee without abstractions. \bsq{Absts. (flat)} and \bsq{Absts. (no flat)} show the performance of Twee once the abstractions are added. Unlike with partial proof abstractions, there is no time-slicing.
The plot does not include the time to learn the abstractions for a given domain, as this is shared between all problems in the domain.
%

We see that applying just domain abstractions to a problem results in a modest speedup. However, applying both domain abstractions and goal flattening leads to a large speedup, with roughly 25 more problems solved inside of 300 seconds. When comparing only the problems that baseline Twee could solve within 300 seconds, adding domain abstractions with goal flattening roughly halves the runtime. Hence domain abstractions seem to interact particularly well with goal flattening.

We see that partial proof abstractions do provide some improvement in runtimes, and solve more problems after 300s, but are not as effective as domain abstractions. However, they are still useful in many cases since they do not require any pre-processing to find the abstractions. Furthermore, they have the advantage of being available for any problem, even those for which we have no other proofs to learn from.

\subsection{Solving Hard Problems}
\label{sec:hard_problems}
Along with general runtime improvements, we find that adding Twitch abstractions allows us to prove 18 hard problems. The full list of hard problems solved can be found in Appendix~\ref{sec:hard-problems-solved}. We give a summary of the results here, and highlight some interesting cases.

\subsubsection{Partial Proof Abstractions}

Given a hard problem, we run Twee on it for at most 500 seconds. We then generate abstractions from the top lemmas of the partial proof, and supply these to Twee. This allowed us to solve 11 hard problems -- each with a total time limit of 1000 seconds, including the partial proof run. The majority of these (6 out of the 11) are LCL problems that contain translations of Horn clauses into equational logic (using the technique introduced in \cite{claessen2018}). In these cases, we find that partial proof abstractions of the form $ifeq(x, true, y, true)$ are always discovered -- encodings of implication in terms of the $ifeq$ function. And through manual testing, we find that these are the abstractions that make the real difference in helping solve the hard problems.

\subsubsection{Domain Abstractions}
\label{sec:domain_hard_problems_solved}
Domain abstractions solved 18 hard problems within 1000 seconds, including every hard problem solved by partial proof abstractions.
For instance, the problem \texttt{LAT075-1} (Rating: 1.00) concerning a Sheffer stroke axiomatization was solved in 32 seconds (without goal flattening) by providing the following domain abstractions:
\begin{itemize}
\item \ttfamily\small \seqsplit{$f(A, A)$}
\item \ttfamily\small \seqsplit{$f(f(B, A), f(B, A))$}
\item \ttfamily\small \seqsplit{$f(f(A, B), f(A, C))$}
\item \ttfamily\small \seqsplit{$f(f(D, C), f(B, A))$}
\item \ttfamily\small \seqsplit{$f(f(f(C, f(C, f(f(A, A), B))), B), A)$}
\end{itemize}
Checking manually, we found that the first and third abstractions were crucial to finding a proof in a reasonable amount of time.

We note that at least the first three abstractions are semantically meaningful. The first one corresponds to the \emph{not} operator, and the second one to \emph{and}. The Sheffer stroke satisfies a kind of self-distributivity law (in which a \emph{nand} distributes over a \emph{nand} to create a \emph{nor}) and the third time represents a term to which this can be applied.

Further inspection of the full results (supplied in the source repository) shows that the abstraction list are quite long in a few cases (three to be exact). These cases highlight the fact that in certain scenarios, it is helpful to skip abstracting over the good abstractions (line 10 of Procedure~\ref{alg:get_domain_abstractions}) and simply take $A_\mathcal{D}$ to be the set of all good abstractions.

We also note that, in addition to the 18 problems, Problem \texttt{GRP677-1.p} was also solved (in 180 s) using domain abstractions. However, this required constraining the domain $\mathcal{D}$ to only include problems that share very similar axioms to the hard problem being solved. So, we see that a larger domain is not necessarily better. This suggests that, when aiming to solve a particular hard problem, curation of the domain we provide Twitch may be a useful direction for future work.

\subsection{Abstractions vs Definitional Axioms}
\label{sec:ablations}

There are two ways one might use \stitch{} abstractions to augment a problem $P$ -- by giving them to Twee as abstractions, or by adding them as definitional axioms (as described in Section~\ref{sec:example}). Twitch uses the abstractions approach. We here present experiments that aim to validate this choice, by comparing the two approaches. The experiments involved running Algorithm~\ref{alg:local_abstractions} on the TPTP problems listed in Table~\ref{tab:tptpv9}.

\begin{figure}[t]
    \centering
    \begin{subfigure}{0.49\textwidth}
        \centering
        \includegraphics[width=\linewidth]{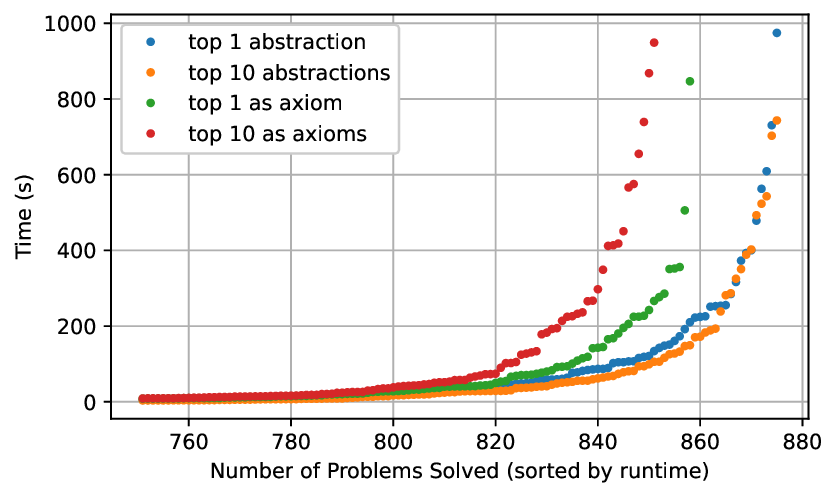}
        \caption{Cactus plot of runtimes for varying number of abstractions and ways of providing them to Twee -- as axioms or abstractions.}
        \label{fig:axvshint_time}
    \end{subfigure}
    \hfill
    \begin{subfigure}{0.49\textwidth}
        \centering
        \includegraphics[width=\linewidth]{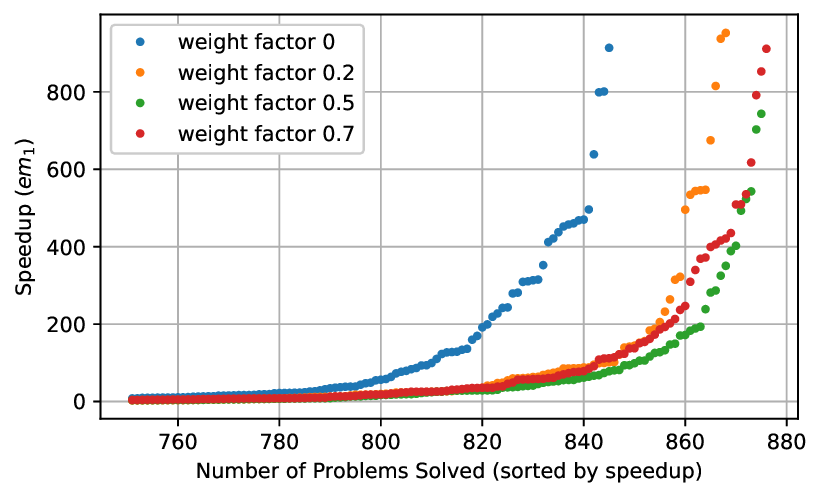}
        \caption{Cactus plot of runtimes for varying weight factors (see Section \ref{sec:hints_in_twee}).}
        \label{fig:hints_skelfactor}
    \end{subfigure}
    \caption{Cactus plots illustrating the effect of local abstractions. A given point in these plots corresponds to a particular problem $P$ from Table \ref{tab:tptpv9}. The y-value of a point corresponds to the time taken by Twee to find a proof of $P$ when augmented with local abstractions, i.e. abstractions derived from an existing proof of $P$ by default Twee. The x-value corresponds to the index of the problem when the problems are sorted by increasing runtime of Twee $P$ augmented with the local abstractions. In other words, the points correspond to runtimes from line \ref{line:augmented_run} of Algorithm~\ref{alg:local_abstractions}.}
\end{figure}

Figure~\ref{fig:axvshint_time} depicts cactus plots of runtimes obtained in line \ref{line:augmented_run} of Algorithm~\ref{alg:local_abstractions} for each problem. The different colours correspond to different approaches in augmenting $P$ with $A$. The \bsq{top $k$} in the figure legends refers to taking the top $k$ abstractions found by \stitch{} in line \ref{line:run_stitch} of Algorithm~\ref{alg:local_abstractions}. \footnote{Note that we do not necessarily actually use all $k$ abstractions to augment the problem, as some may not be back-translatable (see translation discussion in Section \ref{sec:translation}). Nonetheless, the comparison remains informative as the actual number of abstractions used for a given problem during this experiment was the same for both the abstraction and axiom approaches.}.

We see that overall abstractions lead to lower runtimes compared to definitional axioms, and in particular, fewer cases in which Twee times out as a result of the augmentation.
As such, the abstractions approach provides a larger pool of good abstractions for Twitch to derive domain abstractions from. Furthermore, we see that increasing from 1 to 10 abstractions leads to a significant increase in timeouts when using axioms, while the abstractions approach is much more robust to this increase. Hence, using the domain abstractions derived from Twitch as Twee abstractions to augment some hard problem allows us to provide more abstractions with a lesser chance of incurring substantial slowdowns.

However, there are still some problems where providing abstractions as axioms lead to more significant speedups compared to using Twee's abstraction mechanism. For example, our earlier motivating \texttt{LAT075-1} problem illustrates such a case. \footnote{There are also a few problems (15 to be exact) where axioms lead to speedups while abstractions lead to slowdowns. With $P$ taken as LAT141-1, we get a proof $\Gamma(P)$ from default Twee (without goal-flattening) in $\sim 160$ seconds. The top abstraction found by \stitch{} on this proof is $f(X, Y, Z) = meet(X, join(Y, Z))$. Adding this as an axiom leads to a proof of $P \augment A$ in $\sim 12$ seconds, while with the abstraction mechanism, it leads to a proof in $\sim 260$ seconds.}
This suggests that there may be room for combining the two approaches in future work.

Finally, we include Figure~\ref{fig:hints_skelfactor} which shows a cactus plot of abstraction-augmented runtimes with varying weight factors. A weight factor of about 0.5 (halving the cost of terms matching abbreviations) seems to work best. This aligns with the intuition that low weight factors are more aggressive and hence lead to more slowdowns (in particular, timeouts). However, some individual problems have the greatest speed-ups with lower weight factors.


\section{Related Work}

Our implementation of abstractions in Twee is an automatically controlled instance of the \emph{weighting strategy} \cite{weighting}. Weighting later influenced the work of Wos on the \emph{resonance strategy} \cite{wos1995}. Resonators are clause patterns that bias clause selection toward shapes deemed promising by a human expert. This approach proved effective in difficult domains \cite{beeson2016,fitelson2001}. At around the same time Veroff introduced the \emph{hint strategy} \cite{veroff1996} as well as the \emph{proof sketches} method \cite{veroff2001}, in which a prover first establishes an easier or related theorem and then reuses clauses from that proof as hints to guide the search for the harder goal. Our abstractions are in spirit the opposite of Veroff's hints: hint clauses are key clauses which might be derived only once, whereas our abstractions are key term structures which appear over and over again.

Closely related to (and inspiration for) our use of \stitch{} is previous work on automated proof compression. Urban et al.~\cite{vyskovcil2010} propose introducing new definitions to shorten proofs found by saturation provers. Their goal is proof refactoring and compression, whereas we use compression primarily as a mechanism for discovering search-guiding patterns. Similarly, recent work by Wernhard and Zombori \cite{wernhard2025} explores grammar-based compression of Metamath proofs.

Related to the idea that internal proof structure can be mined to improve future search is work on learning-based guidance for saturation provers, such as ENIGMA \cite{urban2017,enigmang2019}. However, these approaches rely on statistical models for clause selection, whereas our method extracts explicit syntactic abstractions that provide lightweight and interpretable guidance tailored to equational completion.

The automatic ranking of interesting derived results has been studied by Puzis et al.~\cite{puzis2006}, who use automated reasoning to generate and filter potentially interesting theorems. Their work inspired our simple heuristic for ranking lemmas extracted from partial Twee proofs, and further exploration of their measures in our setting is an interesting direction for future work.

Other approaches to learning from related problems have been explored in automated reasoning. For example, the \emph{Octopus} system \cite{newborn2004octopus} combines learning and parallel search, using information from easier problems to guide proofs of harder ones. This aligns with our domain abstraction approach, which is inspired by curriculum learning \cite{curriculum-learning}: simpler problems in a domain provide reusable structural patterns for harder ones.

Automated theorem provers are increasingly used as genuine research tools in algebra to obtain mathematically interesting results. For instance, many results in quasigroup and loop theory have been obtained with the assistance of Prover9 and Otter\cite{kinyon2005,phillips2010}. In such settings, the usefulness of ATPs depends not only on finding proofs, but also on producing derivations that can be understood and reformulated in standard mathematical terms. Techniques such as proof sketches and proof simplification address this need. Our abstractions are related in spirit: the recurring term patterns identified by Twitch would ideally correspond to algebraically meaningful constructions, and guiding proof search toward these patterns may help produce proofs that are more structured and easier to analyse.





\section{Limitations and Future Work}

The system we have presented is in many ways quite simple-minded, and there are many ways in which it can be refined. Firstly, our domain construction is quite crude: we restrict to problems within the same TPTP theory sharing concrete symbols, and we do not attempt to automatically generate related or weakened conjectures in the style of proof sketches. Likewise, lemma scoring and term extraction from proofs are based on simple heuristics that could likely be refined as a path to more useful abstractions.

Further, abstraction selection is treated as a single-objective optimization problem focused primarily on runtime speedup. Alternative or multi-objective criteria — such as proof length reduction or robustness across strategies — may yield better guidance. It may also be fruitful to learn abstraction proposals directly from accumulated (problem, abstraction) training data, potentially via language models. Despite these simplifications, we believe our results demonstrate that even a relatively simple abstraction pipeline can produce abstractions yielding substantial improvements and solve previously intractable problems, and that this suggests that automated, structure-based abstraction discovery is a promising direction for future research in equational theorem proving.
We also note that for some problems, the best runtimes were obtained by providing both domain and partial proof abstractions. In particular, one more hard problem -- \texttt{LCL351-10} -- was solved this way, where all other approaches failed, indicating potential of further benefits through such combination.


Also, the implementation of abstractions in Twee is currently quite brittle. The problem is that we only give bonuses to critical pairs that contain instances of the abstraction terms. However, Twee (in common with superposition provers) demodulates critical pairs, simplifying them using the generated rewrite rules. Thus it can easily happen that a term matching an abstraction gets rewritten to a term that does not match. For example, consider the abstraction $f(x,x)$. If $f$ is also associative ($f(f(x,y),z) \rightarrow f(x,f(y,z))$) then the term $f(f(x,x),y)$ (which matches the abstraction) will be rewritten to $f(x,f(x,y))$ (which does not). We are experimenting with a critical pair-like mechanism for generating extra \emph{derived abstractions} when an abstraction overlaps with a rewrite rule in such a way that rewriting destroys the abstraction. We are able to generate many extra abstractions this way, but the generated abstractions can very different from the user-provided abstractions and it is not clear if they respect the user's intent. 


Finally, we would like to explore generating not only abstractions but also hints. These are key steps that are needed in a proof, but may be difficult for the prover to find. An example of a good hint might be a rewrite rule which is large (hence is unlikely to be selected by Twee) but leads to a proof of a conjecture or important lemma.

\section{Conclusion}

We have shown that abstractions, automatically generated via compression of existing proofs, are a promising strategy for equational proving. Augmenting Twee with pre-learned domain abstractions allows us to prove several difficult conjectures and roughly halves the prover's runtime for more typical conjectures.

Creating high-quality abstractions currently requires a supply of simpler problems within the same domain. In cases where these do not exist, we can also generate abstractions from a failed proof attempt, which are currently lower quality. Since typically only a few simple domain abstractions are needed, we hope that a more sophisticated extraction mechanism can solve this problem.

There is clearly much potential to refine the method, both how the abstractions are found and how they are exploited by the prover. Even so, we believe the results are promising and motivate further exploration of this idea.

\begin{credits}
\subsubsection{\ackname} This work was partially supported by the Wallenberg AI, Autonomous Systems and
Software Program (WASP) funded by the Knut and Alice Wallenberg Foundation, and by the Swedish Research Council (VR) grant 2025-06153, Semantically-guided theorem proving for mathematics.

\end{credits}
%
%
%
 \bibliographystyle{splncs04}
 \bibliography{refs}

\appendix

 \section{Experiment Parameters}
 \label{sec:parameters}
There are various parameters that influenced the performance of Twitch and Twee when running the experiments. We list them here.
General \stitch{} and Twee parameters:
\begin{itemize}
    \item The number of abstractions \stitch{} is allowed to return.
    \item The maximum arity of the abstractions returned by \stitch.
    \item Whether to run Twee with $(\texttt{flattening} = True)$ or without $(\texttt{flattening} = False)$ goal flattening.
    \item The \texttt{weight-factor} parameter of Twee influencing the effect of abstractions.
\end{itemize}
Parameters specific to domain abstractions:
\begin{itemize}
    \item The strategies to use during the local abstraction phase.
    \item The timelimit for Twee runs during the local abstraction phase.
    \item The threshold $\tau$ for determining good abstractions (see line 9 of Algorithm~\ref{alg:get_domain_abstractions}).
    \item Whether to run the final \stitch{} step in line 10 of Algorithm~\ref{alg:get_domain_abstractions} to compress the good abstractions; or simply take all good abstractions as domain abstractions.
\end{itemize}
Parameters specific to partial proof abstractions:
\begin{itemize}
    \item The amount of time $t_{\text{par}}$ to run Twee for to generate the partial proof.
    \item The number $k$ of top lemmas to take from the partial proof.
    
\end{itemize}

Throughout the experiments, the \stitch{} maximum arity parameter is fixed with the value 5. Similarly, the number of abstractions \stitch{} is allowed to return is fixed at 10.\\

To produce the domain abstractions used throughout the paper, the \textsc{LocalAbstractions} phase consisted of running each domain problem (for 1000 seconds) for a fixed collection of strategies. If we identify a strategy with a pair $(\texttt{flattening}, \texttt{weight-factor})$, then the set of strategies we used is the product $\{True, False\} \times \{0, 0.2, 0.5, 0.7\}$.\\

In Figure \ref{fig:partial-timeslice}, we take the top $50$ partial proof lemmas and have $\texttt{weight-factor} = 0.2$.\\
In Figure \ref{fig:domain-timeslice}, we have $\texttt{weight-factor} = 0.2$, $\tau = 0.2$, and the final \stitch{} step enabled.\\

To produce the results on the hard problems solved with partial proof abstractions, we ran Twee (with goal flattening set to \texttt{flattening}) for $t_{\text{par}}$ seconds to generate the partial proof, and then took the top $k$ lemmas from that proof to generate abstractions from. We then reran Twee (with goal flattening set to \texttt{flattening}, and $\texttt{weight-factor} = 0.2$) on the hard problem augmented with the partial proof abstractions for the remaining $(1000 - t_{\text{par}})$ seconds. We did this for all $(\texttt{flattening}, t_{\text{par}}, k) \in \{True, False\} \times \{50, 150, 500\} \times \{50, 100\}$.\\

To produce the results on the hard problems solved with domain abstractions, we iterated over a spread of 8 combinations of the relevant parameter values, which each have varying success on different problems. Specifically, we tested the following parameter combinations for each problem:
\begin{itemize}
    \item $\texttt{flattening} = True$; $\texttt{abstraction-weight-factor} = 0.2$; $\tau = 0.7$; final \stitch{} step enabled.
    \item $\texttt{flattening} = False $; $\texttt{abstraction-weight-factor} = 0$; $\tau = 0.2$; final \stitch{} step disabled.
    \item $\texttt{flattening} = False $; $\texttt{abstraction-weight-factor} = 0$; $\tau = 0.7$; final \stitch{} step disabled.
    \item $\texttt{flattening} = True$; $\texttt{abstraction-weight-factor} = 0$; $\tau = 0.5$; final \stitch{} step disabled.
    \item $\texttt{flattening} = False $; $\texttt{abstraction-weight-factor} = 0.5$; $\tau = 0.5$; final \stitch{} step enabled.
    \item $\texttt{flattening} = False $; $\texttt{abstraction-weight-factor} = 0.2$; $\tau = 0.2$; final \stitch{} step enabled.
    \item $\texttt{flattening} = False $; $\texttt{abstraction-weight-factor} = 0.5$; $\tau = 0.5$; final \stitch{} step enabled.
    \item $\texttt{flattening} = True$; $\texttt{abstraction-weight-factor} = 0.2$; $\tau = 0.2$; final \stitch{} step disabled.
\end{itemize}
Here, $\texttt{flattening}$ refers to the goal flattening parameter of Twee when run on the hard problem augmented with the domain abstractions.\\

In Figures \ref{fig:axvshint_time} and \ref{fig:hints_skelfactor}, all runs were done with $\texttt{flattening} = False$. In Figure \ref{fig:axvshint_time}, the \texttt{abstraction-weight-factor} is set to $0.5$ (this has no effect on the definitional axiom runs).

\section{Hardware}

All experiments were run on an AMD Zen 4 EPYC 9354 core.

 \section{Hard Problems Solved}
 \label{sec:hard-problems-solved}

The following hard problems were solved with the help of abstractions derived by Twitch. Recall that the ratings are those as of TPTP v9.2.1. For the full list of successful runs and their associated hints and experiment parameter values, see the \texttt{json} files in the \texttt{data/experiments/hard\_successes} directory of the code repository: \url{https://github.com/WeAreDevo/ijcar26-twee_abstractions}.

\subsection{Partial Proof Abstractions}
\label{sec:partial_abs}

Using partial proof abstractions:
\begin{itemize}
  \item \texttt{GRP723-1.p} (Rating: 1.00) -- Best time: 418.47s
  \item \texttt{LAT075-1.p} (Rating: 1.00) -- Best time: 141.11s
  \item \texttt{LCL375-10.p} (Rating: 1.00) -- Best time: 399.95s
  \item \texttt{LCL395-10.p} (Rating: 1.00) -- Best time: 376.37s
  \item \texttt{LCL225-10.p} (Rating: 0.96) -- Best time: 11.55s
  \item \texttt{LCL245-10.p} (Rating: 0.96) -- Best time: 114.44s
  \item \texttt{LCL289-10.p} (Rating: 0.96) -- Best time: 130.55s
  \item \texttt{LCL348-10.p} (Rating: 0.96) -- Best time: 11.56s
  \item \texttt{ROB031-2.p} (Rating: 0.96) -- Best time: 12.78s
  \item \texttt{ROB035-1.p} (Rating: 0.96) -- Best time: 12.30s
  \item \texttt{LAT400-1.p} (Rating: 0.91) -- Best time: 698.88s
\end{itemize}

\subsection{Domain Abstractions}
\label{sec:domain_abs_list}

Using domain abstractions:

\begin{itemize}
  \item \texttt{GRP672-11.p} (Rating: 1.00) -- Best time: 510.31s
  \item \texttt{GRP723-1.p} (Rating: 1.00) -- Best time: 708.57s
  \item \texttt{GRP732-1.p} (Rating: 1.00) -- Best time: 549.19s
  \item \texttt{LAT075-1.p} (Rating: 1.00) -- Best time: 31.94s
  \item \texttt{LCL304-10.p} (Rating: 1.00) -- Best time: 119.69s
  \item \texttt{LCL331-10.p} (Rating: 1.00) -- Best time: 40.88s
  \item \texttt{LCL375-10.p} (Rating: 1.00) -- Best time: 119.14s
  \item \texttt{LCL395-10.p} (Rating: 1.00) -- Best time: 522.33s
  \item \texttt{ROB032-2.p} (Rating: 1.00) -- Best time: 41.19s
  \item \texttt{ROB034-1.p} (Rating: 1.00) -- Best time: 42.35s
  \item \texttt{LCL225-10.p} (Rating: 0.96) -- Best time: 295.57s
  \item \texttt{LCL245-10.p} (Rating: 0.96) -- Best time: 125.09s
  \item \texttt{LCL289-10.p} (Rating: 0.96) -- Best time: 105.07s
  \item \texttt{LCL339-10.p} (Rating: 0.96) -- Best time: 247.18s
  \item \texttt{LCL348-10.p} (Rating: 0.96) -- Best time: 31.57s
  \item \texttt{ROB031-2.p} (Rating: 0.96) -- Best time: 33.79s
  \item \texttt{ROB035-1.p} (Rating: 0.96) -- Best time: 32.73s
  \item \texttt{LAT400-1.p} (Rating: 0.91) -- Best time: 160.89s
\end{itemize}

Using a hand-picked set of domain problems:

\begin{itemize}
  \item \texttt{GRP677-1.p} (Rating: 1.00) -- Best time: 180s
\end{itemize}
Using domain abstractions together with partial proofs:

\begin{itemize}
  \item \texttt{LCL351-10.p} (Rating: 1.00) -- Best time: 200s
\end{itemize}



\end{document}